\documentclass[conference]{IEEEtran}
\IEEEoverridecommandlockouts
\usepackage{cite}
\usepackage{amsmath,amssymb,amsfonts}
\usepackage{algorithmic}
\usepackage{graphicx}
\usepackage{textcomp}
\usepackage{xcolor}
\usepackage[ruled,vlined]{algorithm2e}
\def\BibTeX{{\rm B\kern-.05em{\sc i\kern-.025em b}\kern-.08em
    T\kern-.1667em\lower.7ex\hbox{E}\kern-.125emX}}

\def\bb{\mathbf{b}}
\def\bn{\mathbf{n}}
\def\bq{\mathbf{q}}
\def\bb{\mathbf{b}}

\def\bx{\mathbf{x}}
\def\by{\mathbf{y}}
\def\bz{\mathbf{z}}
\def\bn{\mathbf{n}}

\begin{document}

\title{Turbo Autoencoder with a Trainable Interleaver\\
\thanks{This work is supported by Samsung Research America (SMI Lab).}
}

\author{Karl Chahine$^{*}$, Yihan Jiang$^\dagger$, Pooja Nuti$^{*}$, Hyeji Kim$^{*}$, and Joonyoung Cho$^{\ddagger}$ \\
    $^{*}$ University of Texas at Austin, 
    $^{\dagger}$ Aira Technology, 
    $^{\ddagger}$ Samsung Research America}

\maketitle

\begin{abstract} 
A critical aspect of reliable communication involves the design of codes that allow transmissions to be robustly and computationally efficiently decoded under noisy conditions. Advances in the design of reliable codes have been driven by coding theory and have been sporadic. 
Recently, it is shown that channel codes that are comparable to modern codes can be learned solely via deep learning. In particular, Turbo Autoencoder (\textsc{TurboAE}), introduced by Jiang et al., is shown to achieve the reliability of Turbo codes for Additive White Gaussian Noise channels. 
%

In this paper, we focus on applying the idea of \textsc{TurboAE} to various practical channels, such as fading channels and chirp noise channels. We introduce \textsc{TurboAE-TI}, a novel neural architecture that combines \textsc{TurboAE} with a trainable interleaver design. We develop a carefully-designed training procedure and a novel interleaver penalty function that are crucial in learning the interleaver and \textsc{TurboAE} jointly.  
We demonstrate that \textsc{TurboAE-TI} outperforms \textsc{TurboAE} and LTE Turbo codes for several channels of interest. We also provide interpretation analysis to better understand \textsc{TurboAE-TI}. 

\end{abstract}

\begin{IEEEkeywords}
neural channel coding, Turbo autoencoder, deep learning, CNN, fading channels, chirp signal, bursty noise
\end{IEEEkeywords}

\section{Introduction}

The discipline of coding theory has made significant progress in the past seven decades since Shannon's celebrated work~\cite{shannon1948mathematical}. Several codes have been invented for reliable communications, featuring convolutional codes, turbo codes, Low-Density Parity Check (LDPC) codes, and polar codes. 

\begin{figure}[!ht]
    \centering
    \includegraphics[width=0.9\linewidth]{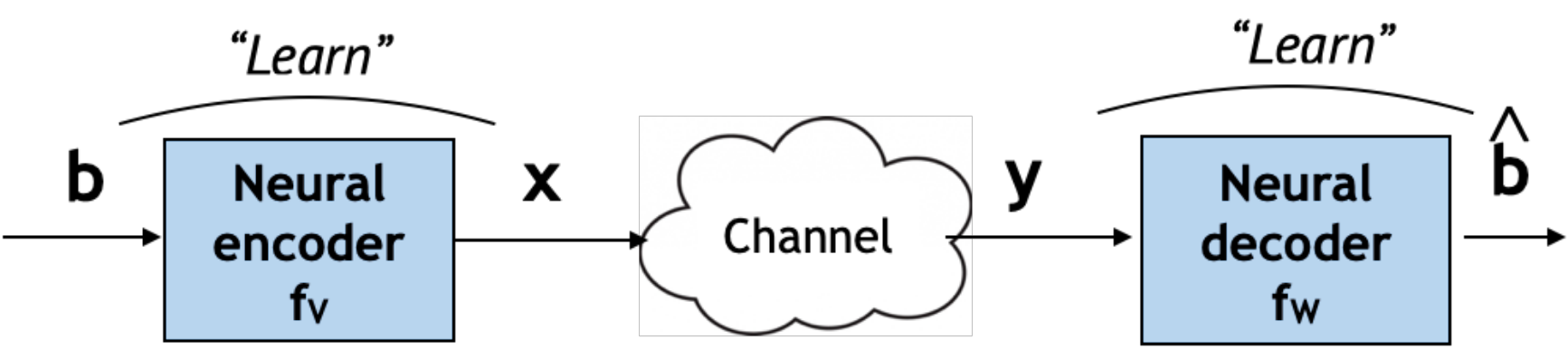}
    \caption{Autoencoder framework for channel coding: channel encoder and decoder are modeled as neural networks and are trained jointly. 
    }
    \vspace{-1em} 
    \label{fig:fig1}
\end{figure}

Deep learning is an emerging and powerful tool that has demonstrated promising success in channel coding~\cite{DBLP:journals/corr/OSheaH17, dorner2017deep, gruber2017deep} as well as other aspects of physical layer communications, such as beamforming~\cite{alkhateeb2018deep}, modulation~\cite{modulation1}, channel estimation~\cite{ye2017power} and channel feedback~\cite{csinet2}. In channel coding, on the one hand, it is shown that introducing learnable components to the existing codes and decoders leads to the improvement in reliability and complexity (e.g., weighted belief propagation decoder for linear codes~\cite{nachmani2018deep,gruber2017deep} and~\cite{shlezinger2019viterbinet,turbonet20}). On the other hand, it is shown that 
end-to-end neural network based codes can perform comparably to the state-of-the-art codes for Additive White Gaussian Noise (AWGN) channels~\cite{Jiang19,kim2018deepcode}. As depicted in Figure~\ref{fig:fig1}, a pair of channel encoder and decoder modeled as neural networks and jointly trained via backpropagation achieves the reliability of modern codes. 

\begin{figure}[!ht]
\centering
\includegraphics[width=0.98\linewidth]{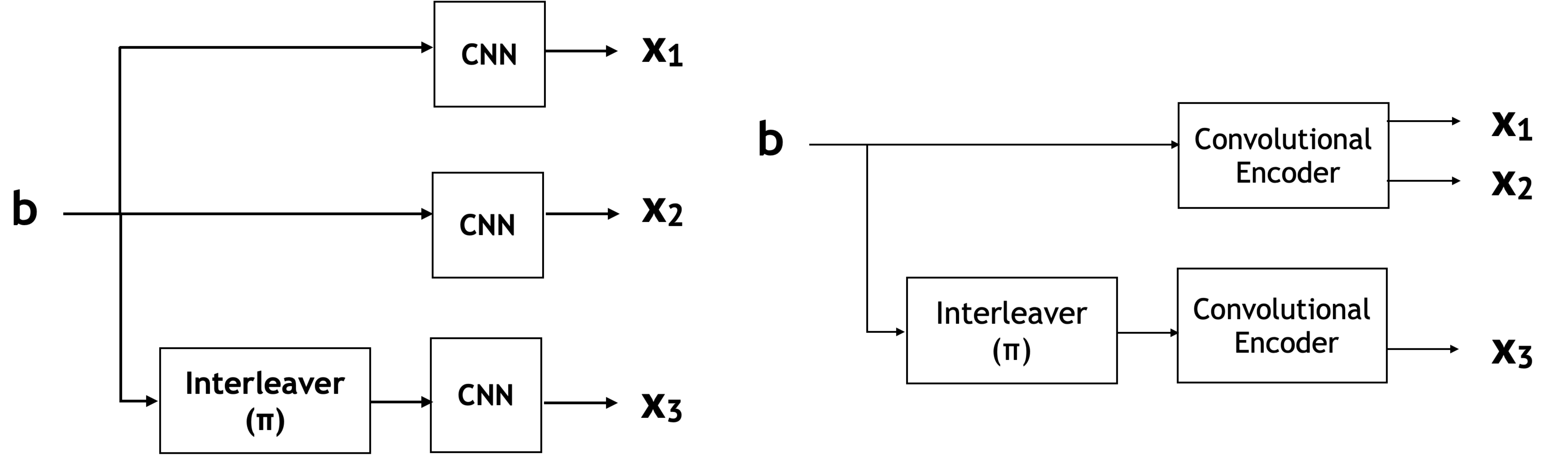} 
\caption{Encoder of \textsc{TurboAE} vs. Turbo codes for a rate-1/3 code. }
\label{fig1}
\end{figure}

While the training framework in Figure~\ref{fig:fig1} might look straightforward, training a reliable code is shown to be highly challenging~\cite{Jiang19}. The generic neural architectures, such as Convolutional Neural Networks (CNN) and Recurrent Neural Networks (RNN), are empirically shown to perform poorly~\cite{Jiang19} as they tend to learn codes with a very short memory. To address these challenges,~\cite{Jiang19} introduces a 
custom neural architecture called \textsc{TurboAE}, which concatenates parity bits generated from three CNNs; as depicted in Figure~\ref{fig1} (left), one of the CNNs takes an interleaved bit stream. which creates a long-range memory. This architecture is similar to Turbo codes, shown in Figure~\ref{fig1} (right). \textsc{TurboAE}, trained via a carefully designed training methodology, mimics the reliability of Turbo codes for AWGN channels~\cite{Jiang19}. 
%

%

Since~\cite{Jiang19}, \textsc{TurboAE} architecture has gained huge interest, and similar approaches for designing channel codes have been proposed. Neural codes inspired by serially concatenated Turbo codes and Reed-Muller/Polar codes are introduced in~\cite{serialturboae} and in~\cite{pmlr-v139-makkuva21a}, respectively. \textsc{TurboAE} has been applied to channels with feedback~\cite{ftae2020} and joint coding and modulation~\cite{turboaemod}. Methods to speed up training are presented in~\cite{serialturboae}. 

Despite the success, there are two important open problems. 
First, how  \textsc{TurboAE} performs on various {\em practical} non-AWGN channels, such as fading channels and channels with bursty interference, is vastly unknown. 
Can we leverage the {\em trainability} of \textsc{TurboAE} for 
challenging communications? 
%
%
%

Second, can we improve \textsc{TurboAE} even further? In particular, a {\em pseudo-random} interleaver is deployed in the encoder of \textsc{TurboAE} in~\cite{Jiang19,serialturboae}. Is this an optimal choice? Does interleaver play an important role in the reliability? If so, how can we choose an interleaver that leads to the high reliability? These are challenging problems as the space of interleaver grows faster than exponential in the block length. 
%
%
%
%
%
%
%
%
%
%
%
%
In this paper, we make progress on both aforementioned problems. 
%
%
Our main contributions are as follows. 

\begin{itemize} 
    \item {\em Development of Turbo Autoencoder with a Trainable Interleaver} (\textsc{TurboAE-TI}): We introduce a penalty function to make the interleaver {\em trainable} and provide a well-thought-out optimization methodology to {\em learn} the interleaver together with the rest of \textsc{TurboAE}. 
    %

    \item {\em Adaptivity and robustness}:
    We demonstrate that \textsc{TurboAE-TI} outperforms \textsc{TurboAE} and LTE Turbo codes on several channels of practical interest, such as  fading channels and bursty noise channels. %
    We show that we gain in reliability up to 1dB.  
    \item {\em Performance on AWGN  channels}: We show that TurboAE-TI outperforms TurboAE and LTE turbo codes for the classical AWGN channels. We develop a novel training methodology, called {\em Rician training}, which is crucial in achieving the high reliability. 
    \item {\em Interpretation}: We provide analysis to interpret the behavior of learned codes and interleavers. Our analysis implies that \textsc{TurboAE-TI} has learned a sophisticated pattern that is beyond maximizing the codeword distance.  
\end{itemize}


In the rest of this paper, we refer to \textsc{TurboAE} by \textsc{TurboAE-UI} (TurboAE with a Uniform Interleaver), to be clear about their interleaver design, which is generated uniformly at random.

\section{Background: Turbo Autoencoder}

In this section, we review Turbo Autoencoder (\textsc{TurboAE-UI}), one of the state-of-the-art neural network-based channel codes introduced in~\cite{Jiang19}. \textsc{TurboAE-UI} 
is inspired by Turbo codes, which concatenates a convolutional code and another convolutional code generated from the {\em interleaved} bit stream. 



\noindent \textbf{Encoder}. 
%
%
As shown in Figure~\ref{fig1}, the neural encoder of \textsc{TurboAE-UI} 
combines the interleaver and convolutional neural networks (CNNs); it consists of three trainable CNN encoding  blocks followed by a power normalization layer. 
The input to the first two CNN encoding blocks is the original bit sequence $\bb$, while the input to the last CNN block is an interleaved bit sequence $\pi(\bb)$. Each of the 3 $\textnormal{CNN}$ blocks consists of 2 layers 1-D CNN followed by a linear layer. We use 100 filters, a kernel of size 5, and a stride of 2. The output $\bx$ of each $\textnormal{CNN}$ encoder block is of shape (L,1).

\noindent \textbf{Interleaver}. One of the key components of \textsc{TurboAE-UI} encoder is the interleaver $\pi$. ~\cite{Jiang19,serialturboae} deploy a {\em pseudo-random} interleaver, generated uniformly at random. In Section~\ref{sec:ti}, we empirically show that this selection is {\em sub-optimal}. 

\begin{figure}[!ht]
\centering
\includegraphics[width=0.99\linewidth]{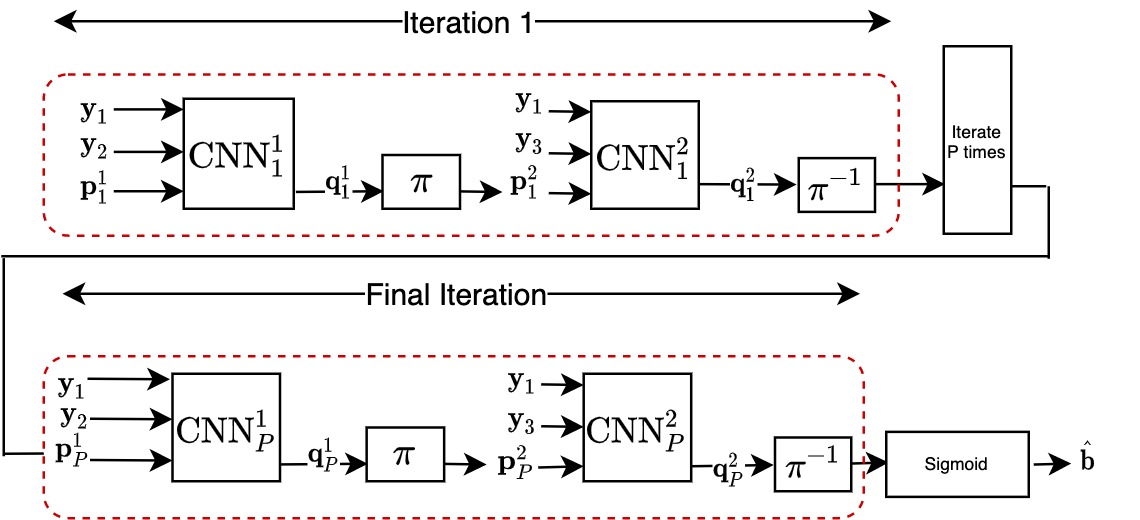} 


\caption{The decoder architecture in Turbo Autoencoder for a rate-1/3 code. }
\label{fig:Turboae-decoder}
\end{figure}

\noindent \textbf{Decoder}. The neural decoder of \textsc{TurboAE-UI} is depicted in  Figure~\ref{fig:Turboae-decoder}. It 
consists of several layers of CNNs with interleavers and de-interleavers in between. 
As received codewords are encoded from original message $\bb$ and interleaved message $\pi(\bb)$, decoding interleaved code requires iterative decoding on both interleaved and de-interleaved order shown in Figure \ref{fig:Turboae-decoder}. 
Let $\by_1,\by_2,\by_3 \in \mathbb{R}^L$ denote noisy versions of $\bx_1, \bx_2, \bx_3 \in \mathbb{R}^L$, respectively. The decoder generates an estimate $\hat{\bb} \in \{0,1\}^L$ based on $\by_1,\by_2,\by_3$ via multiple iterations of CNNs. Two types of decoders are alternatively applied for $P$ iterations, $\textnormal{CNN}^j_i$ for $j  \in \{1, 2\}$ and  $i \in \{1, 2, \cdots , P\}$, with  an interleaver and de-interleaver in between. Each $\textnormal{CNN}^j_i$ block consists of 5 layers 1-D CNN followed by a linear layer. We use 100 filters, a kernel of size 5, and a stride of 2. For $i \in \{1,2,\cdot \cdot \cdot , 11\}$, the output posteriors $\bq^{1}_{i}$ and $\bq^{2}_{i}$ of each decoder block $\textnormal{CNN}^{j}_{i}$ are of shape (L,5), and for $i=12$, the output posterior $\bq^{2}_{P}$ is of shape (L,1). The last iteration generates $\hat{\bb} = sigmoid(\pi^{-1}(\bq^{2}_{P}))$. 

\section{Problem Setup}\label{sec:system-model}

We design channel codes and their corresponding decoders via deep learning. As illustrated in Figure~\ref{fig:fig1}, we model the channel encoder and channel decoder as neural networks and train them jointly via an autoencoder training. For concreteness, we focus on rate-1/3 codes with $L = 40$ information bits (We consider $L=100$ in Section~\ref{sec:open}). 
We aim to minimize the Bit Error Rate (BER), defined as BER $ = \frac{1}{L}\sum_{l=1}^{L}P(\hat{\bb}_{l} \neq \bb_{l})$.
We consider various fading channels and bursty noise channels. 
In particular, we focus on the following models: 
\begin{itemize}

    \item\textbf{Rician} ($K$,$\sigma$): A general description of a Rician fading channel is considered in which a channel is comprised of both a line-of-sight (LOS) and non line-of-sight (NLOS) component. 
    The amplitude of the LOS and NLOS components is dictated by $K$. The channel is defined as  $y_l = h_l x_l + z_l$, where $z_l \sim \mathcal{N}(0,\sigma^2)$ is Gaussian, and 
        \begin{equation}
             h_l = \left|\sqrt{\frac{K}{(K+1)}}(1+1\text{i}) + \sqrt{\frac{1}{(K+1)}}h_l^\text{NLOS}\right|,
        \nonumber
        \end{equation}
        and $h_l^\text{NLOS}$ is distributed as $ \mathcal{CN}(0,I)$.

    \item \textbf{Rician with bursty noise} ($\sigma$, $\sigma_{b}$):
    In this bursty channel model, we proceed in two successive steps. In step 1, we apply fading and add AWGN noise to our coded bits: $y_l = h_l x_l + z_l$, where $h_l$ and $z_l$ are defined as above. In step two, we generate a bursty sequence $\bn \sim \mathcal{N}(0,\sigma^2_{b}I) \in  \mathbb{R}^S$, where $S < L$.
    More specifically, we uniformly pick an index $j \in \{0, 1, \cdot \cdot \cdot, 3L-S-1\}$ and add bursty noise to our noisy sequence: $\by[j:j+S] = \by[j:j+S] + \bn$.
    
    \item \textbf{Rician with linear chirp noise} ($\sigma_{b}$, $f_{0}$, $f_{1}$, $T$, $\phi_{0}$): We also evaluate our model on the linear chirp channel, which consists of applying a high-power frequency-increasing sinusoid to our transmitted bits. The jamming signal in function of time $t$ is given by $q(t) = \sigma^2_{b}[sin(\phi_{0}) + 2\pi(\frac{c}{2}t^2 + f_{0}t)]$, where $\phi_{0}$ is the initial phase (at time $t = 0$) and $c = \frac{f_{1}-f_{0}}{T}$ is the chip rate. $f_{0}$ is the starting frequency at time $t = 0$ and $T$ is the time is takes to sweep from $f_{0}$ to $f_{1}$. 
    
    Similarly to the previously described channel, we proceed in two successive steps. In step 1, we apply fading and add AWGN noise to our coded bits: $y_l = h_l x_l + z_l$, where $h_l$ and $z_l$ are defined as above. In step two, using $q(t)$, we generate a jamming sequence $\bn \sim \mathcal{N}(0,\sigma^2_{b}I) \in  \mathbb{R}^S$, where $S < L$. More specifically, we uniformly pick an index $j \in \{0, 1, \cdot \cdot \cdot, 3L-S-1\}$ and add the bursty noise to our noisy sequence: $\by[j:j+S] = \by[j:j+S] + \bn$.

\end{itemize}

\section{TurboAE-TI: TurboAE with a trainable interleaver}\label{sec:ti}

We introduce a neural architecture, called \textsc{TurboAE-TI}, which combines {\em interleaver training} and \textsc{TurboAE-UI} architecture. 
{\color{black} Learning an interleaver is challenging as the space of interleaver scales in $L!$, where $L$ denotes the block length. To address this challenge, we introduce a carefully-designed training procedure, as well as an interleaver penalty function, which will be optimized in conjunction with our model's loss function. The details are described in Section IV. A-B}.  
As we show in Section~\ref{sec:result}, the {\em interleaver training} leads to a noticeable improvement upon \textsc{TurboAE-UI}. For concreteness, we focus on rate-1/3 codes with $L = 40$ information bits. 

\subsection{Trainable Interleaver} 

We propose to learn an interleaver via deep learning, hence getting rid of its manual design, resulting in a more efficient matrix shuffle. The interleaver operation can be represented as multiplying the message sequence $\textbf{b}$ by a permutation matrix $T \in [0,1]^{L \times L}$, a square binary matrix with exactly one entry of one in each row and each column and zeros elsewhere. Since it's too costly to enumerate all possibilities, we propose to approach it by adding a matrix generalization of the $l_{1}-l_{2}$ penalty \cite{esser2013method} to our model's loss function during training.
 The proposed interleaver penalty $P(T)$ is given by:

\begin{align*}
    P(T) = \sum_{i=1}^{L}\Bigg[\sum_{j=1}^{L}|t_{ij}| - \left(\sum_{j=1}^{L}t_{ij}^2\right)^{1/2}\Bigg] \\+ \sum_{j=1}^{L}\Bigg[\sum_{i=1}^{L}|t_{ij}| - \left(\sum_{i=1}^{L}t_{ij}^2\right)^{1/2}\Bigg]
\end{align*}

Moreover, to satisfy the permutation matrix's properties, we have to abide by the following conditions:

\vspace{-1em}
\begin{align}
\label{constraints}
    t_{ij} \geq 0, \forall(i,j); \sum_{i=1}^L t_{ij} = 1, \forall j ; \sum_{j=1}^L t_{ij} = 1, \forall i
\end{align}\label{eq1}

The training details are discussed in the following sub-section.

\subsection{Training Methodology}

Let $L(w_{enc},w_{dec},T)$ denote our model's loss function. The training thus minimizes:
\vspace{-.7em}
\begin{align*}
    f = L(w_{enc},w_{dec},T) + \lambda \cdot P(T)
\end{align*}

Where $w_{enc}$, $w_{dec}$ are the encoder's and decoder's weights, respectively, and $\lambda \in \mathbb{R}$ is a regularization constant used to balance the contribution of each component in the loss function. The details are described in Algorithm 1. The main differences from conventional training are as follows:
\begin{itemize}
    \item At every epoch, encoders and decoders are trained separately. The number of iterations for the encoders and decoders are $S_{enc}$ and $T_{Sec}$ respectively. This prevents the training from converging to a local optimum.
    \item Empirically, our results showed that using different training noise levels for the encoders ($\sigma_{enc}$) and decoders ($\sigma_{dec}$) resulted in better performance.
    \item The used batch size was very large (500). This is important to average out the channel noise effects.
    \item To satisfy the three constraints in (1), we first clip the values of T: $T \leftarrow max(T,0)$. Then, we normalize each column of $T$ by dividing each column element by the sum of the entries in this particular column. Finally, we normalize each row of $T$ by dividing each row element by the sum of the entries in this particular row.

\end{itemize}

\begin{algorithm}[!htb]
    \SetAlgoLined
    \DontPrintSemicolon
    \textbf{Inputs}: Number of Epochs E, Encoder Training Steps $S_{enc}$, Decoder Training Steps $S_{dec}$, Encoder Training noise $\sigma_{enc}$, Decoder Training noise $\sigma_{dec}$, Learning Rate $\alpha$, Regularizer $\lambda$\\
    \textbf{Outputs}:  Encoder and Decoder Weights $w_{enc}, w_{dec}$, Interleaver Matrix $T$\\
    \For{$e \leq E$}{
        \For{$k \leq S_{enc}$}{
            Generate examples $\bb$, noise $\bz$ $\sim N(0,\sigma_{enc}^2)$\;
            Compute $f$\ using $\bb$, $\bz$\;
            Generate the gradients $\nabla_{w_{enc}}f$, $\nabla_{T}f$\;
            Update $w_{enc}$: $w_{enc} \leftarrow w_{enc} - \alpha \cdot \nabla_{w_{enc}}f$\;
            Update $T$: $T \leftarrow T - \alpha \cdot \nabla_{T}f$\;
            Non-negativity constraint: $T \leftarrow max(T,0)$\;
            Normalize each row and column of $T$\;

        }

        \For{$k \leq S_{dec}$}{
            Generate examples $\bb$, noise $\bz$ $\sim N(0,\sigma_{dec}^2)$\;
            Compute $f$\ using $\bb$, $\bz$\;
            Generate the gradients $\nabla_{w_{dec}}f$, $\nabla_{T}f$\;
            Update $w_{dec}$: $w_{dec} \leftarrow w_{dec} - \alpha \cdot \nabla_{w_{dec}}f$\;
            Update $T$: $T \leftarrow T - \alpha \cdot \nabla_{T}f$\;
            Non-negativity constraint: $T \leftarrow max(T,0)$\;
            Normalize each row and column of $T$\;

        }
        
    }
    
    \caption{Training of \textsc{TurboAE-TI}}\label{arg1}
\end{algorithm}

\section{Results and analysis} \label{sec:result}

We show that  \textsc{TurboAE-TI} outperforms \textsc{TurboAE-UI} and LTE Turbo codes for fading channels, bursty noise channels, and AWGN channels. We consider various SNRs, defined as SNR $= -10\log_{10}(\sigma^2)$. To measure performance, we plot BER vs $E_{b}/N_{0}$, where $E_{b}$ is the energy per bit, and $N_{0}$ is the noise power. The relation between SNR and $E_{b}/N_{0}$ is given by $E_{b}/N_{0} = 10\log_{10}(\text{SNR}) -10\log_{10}(\text{rate})$. The rate is fixed as 1/3. Interpretation analysis is followed in Section~\ref{sec:int}. 

\subsection{Fading channels}
\begin{figure}[!ht]
    \centering
    \includegraphics[width=0.7\linewidth]{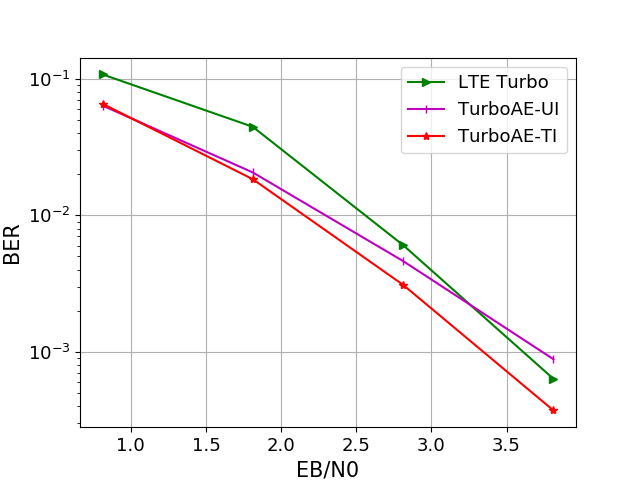}
    \includegraphics[width=0.7\linewidth]{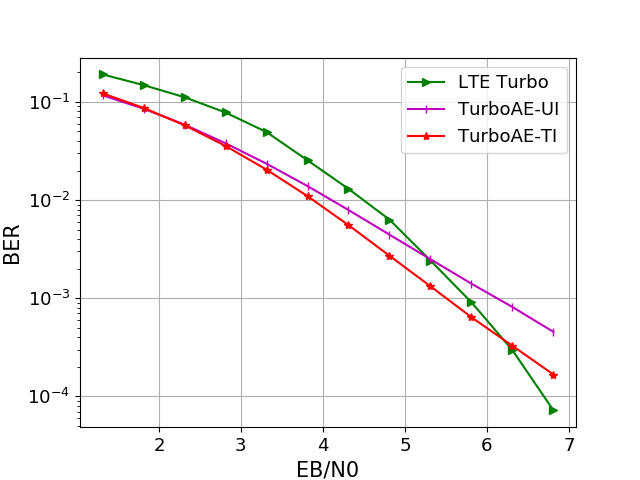}
    \caption{\textsc{TurboAE-TI} outperforms \textsc{TurboAE-UI} and LTE Turbo codes for Rician channels ($K=10$) (top) and Rayleigh channels (bottom) on low-to-moderate $E_b/N_0$ regimes.} 
    \label{fig:fading}
\end{figure}

%
%
We consider Rician channels described in Section~\ref{sec:system-model} (for $K=10$) in Figure~\ref{fig:fading} (top), and Rayleigh channels, for which there is no line of sight component (i.e., Rician channel with $K=0$) in Figure~\ref{fig:fading} (bottom). \textsc{TurboAE-TI}, coupled with a neural interleaver and a personalized training, shows a clearly better performance when compared to \textsc{TurboAE-UI} and LTE Turbo in the low-to-moderate $E_{b}/N_{0}$ regime.


\vspace{-1em}

\vspace{-.2em}
\subsection{Adaptivity and Robustness }
In this subsection, we analyze the \emph{robustness} and \emph{adaptivity} of \textsc{TurboAE-TI} and \textsc{TurboAE-UI}. \emph{Robustness} refers to the ability of a network trained for a particular channel model to work well on a differing channel model without re-training. \emph{Adaptivity} refers to the ability of the network to adapt and retrain for differing channel models with minimal re-training.

\begin{figure}[!ht]
    \centering
    \includegraphics[width=0.7\linewidth]{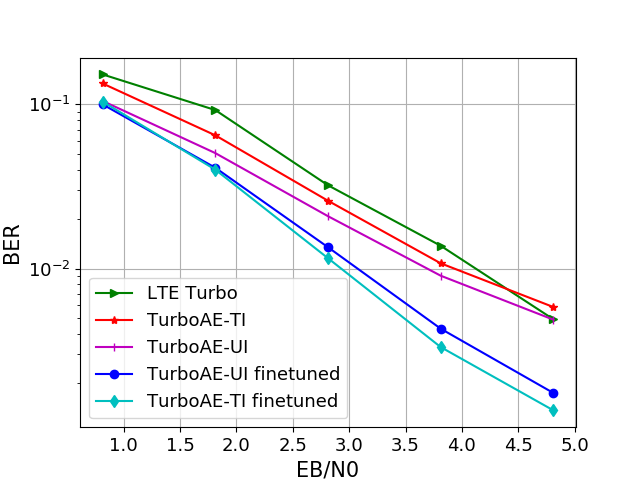}
    \caption{Fine-tuned {TurboAE-TI} outperforms \textsc{TurboAE-UI} and LTE Turbo for Bursty Rician channels ($K=10$) on low-to-moderate $E_b/N_o$ regimes.} 
    \label{fig:bursty_model_2}
\end{figure}

To measure these two metrics, we conduct two experiments. In the first experiment, we consider applying a bursty noise to the Rician channel. The setup is described in Section~\ref{sec:system-model} (Rician with bursty noise). The results are shown in Figure~\ref{fig:bursty_model_2}. First, to measure \emph{robustness}, we consider training \textsc{TurboAE-TI} and \textsc{TurboAE-UI} on Rician channels $(K = 10)$, and testing those models on Rician with bursty noise. Although the models show an improved performance to LTE Turbo, the gap is not significant. Moreover, \textsc{TurboAE-UI} outperforms \textsc{TurboAE-TI}, which is not a satisfactory result. To address those issues, we propose {\em fine-tuning} our models, by training them on Rician channels coupled with bursty noise, for 100 epochs. (\textsc{TurboAE-TI} finetuned and \textsc{TurboAE-UI} finetuned). After {\em fine-tuning} for as little as 100 epochs, ($a$) the performance of \textsc{TurboAE-UI} and \textsc{TurboAE-TI} improves dramatically compared to LTE Turbo, and ($b$) \textsc{TurboAE-TI} performs better than \textsc{TurboAE-UI}. Those results highlight the \emph{adaptivity} of \textsc{TurboAE-TI}.

\begin{figure}[!ht]
    \centering
    \includegraphics[width=0.7\linewidth]{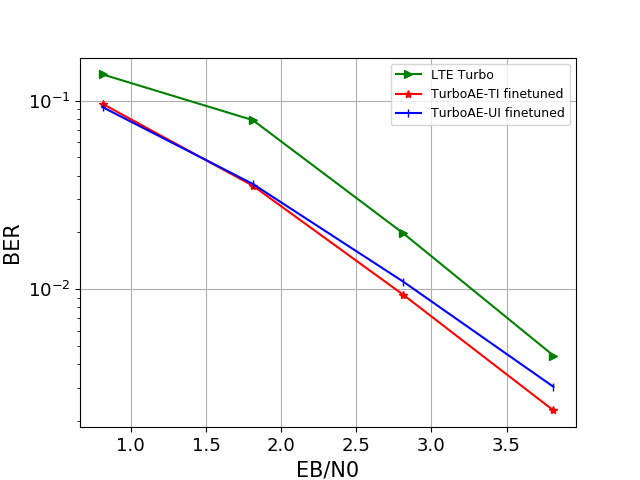}
    \vspace{-.2em}
    \caption{Robustness of \textsc{TurboAE-TI} and \textsc{TurboAE-UI} on 
    Rician fading channels with chirp noise} 
    \label{fig:bursty_chirp}
    \vspace{-.8em}
\end{figure}

In the second experiment, we aim to measure the \emph{robustness} of our models to a different type of bursty noise: chirp jamming signals (described in Section~\ref{sec:system-model}). We use our previously described models (\textsc{TurboAE-TI} finetuned and \textsc{TurboAE-UI} finetuned), and test them on the Rician channel with chirp jamming. The results are shown in Figure~\ref{fig:bursty_chirp}. We notice that both \textsc{TurboAE-TI} and \textsc{TurboAE-UI} show good \emph{robustness} to the modified channel when compared to LTE Turbo, and \textsc{TurboAE-TI} finetuned has a slightly better performance in the high $E_b/N_0$ regime compared to \textsc{TurboAE-UI}.

\vspace{-1em}



\vspace{0em} 
\subsection{AWGN channels}

Given the reliability improvement of \textsc{TurboAE-TI} upon \textsc{TurboAE-UI} on fading and bursty noise channels, a natural question is whether \textsc{TurboAE-TI} outperforms \textsc{TurboAE-UI} on AWGN channels, where 
%
$\mathbf{y} = \mathbf{x} + \mathbf{z}$, where $ \mathbf{z} \sim \mathcal{N}(0,\sigma^2I)$ is the IID Gaussian noise.

\begin{figure}[!htb]
    \centering
    \includegraphics[width=0.67\linewidth]{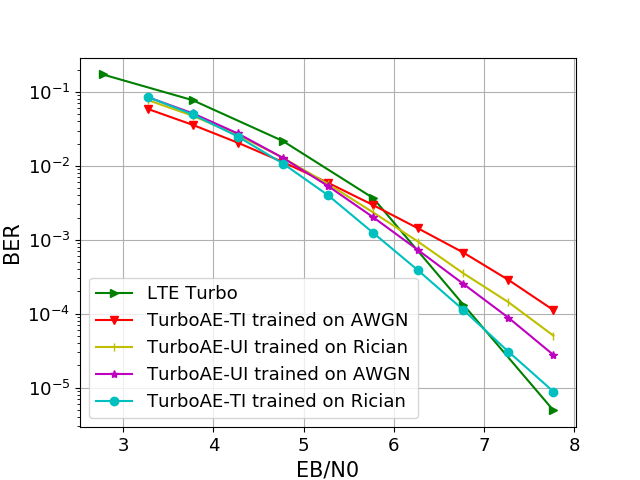}
    \caption{TurboAE-TI trained on Rician fading channels outperforms TurboAE-UI and LTE Turbo codes for AWGN channels. 
    } 
    \label{fig:awgn}
\end{figure}

We train and test both \textsc{TurboAE-TI} and  \textsc{TurboAE-UI} on AWGN channels in Figure~\ref{fig:awgn}, from which we observe that \textsc{TurboAE-TI} (-$\triangledown$-) performs worse than \textsc{TurboAE-UI} (-*-). Furthermore, both perform worse than LTE Turbo codes. We conjecture this is because training an interleaver along with the rest of \textsc{TurboAE-UI} is challenging. 

To mitigate such challenge in training \textsc{TurboAE-TI}, we apply the idea of {\em Rician training} inspired by the promising results on fading channels.  
Instead of training \textsc{TurboAE-TI} on AWGN channels, we train it on Rician fading channels (with $K=10$). As shown in Figure~\ref{fig:awgn}, \textsc{TurboAE-TI} trained for Rician fading channels perform noticeably better than other \textsc{TurboAE-UI} models. 
On the other hand, Rician training does not improve the reliability of \textsc{TurboAE-UI}. 

\vspace{-1.2em}

\vspace{-0.2em}
\section{Interpretation} \label{sec:int}

A natural question is `what has \textsc{TurboAE-TI} learned?'. Although interpreting the behavior of deep learning models is in general challenging, we run several experiments to better understand the behavior of our trained network. The results are described below.

\begin{figure}[!ht]
    \centering
    \includegraphics[width=0.7\linewidth]{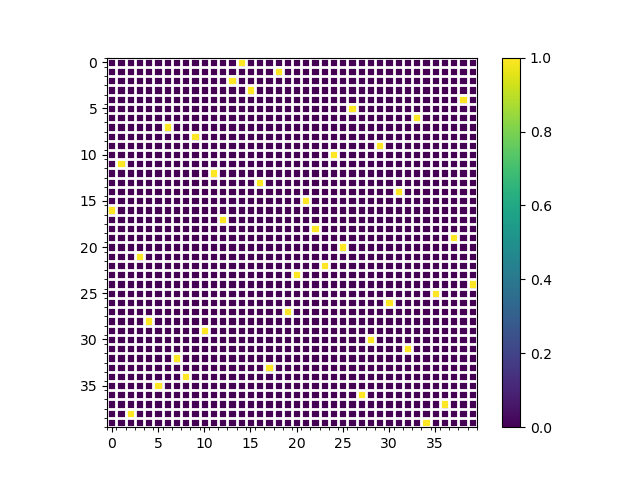}
    \vspace{-1em}
    \caption{Learned Interleaver $T$ for Rician Channel Training 
    }
    \label{fig:Trainable_Interleaver}
    \vspace{-.6em}
\end{figure}

\noindent\textbf{Visualization of the learned interleaver. }\ In Fig. \ref{fig:Trainable_Interleaver}, we visualize the $40 \times 40$ learned permutation matrix $T$ of \textsc{TurboAE-TI} trained on the Rician channel. Yellow (purple) squares denote the positions of 1's (0's). We conclude we learned a legitimate interleaver, given that $T$ satisfies  constraints in \eqref{eq1}.

\vspace{-1.2em}

\begin{figure}[!ht]
\centering
\includegraphics[width=0.7\linewidth]{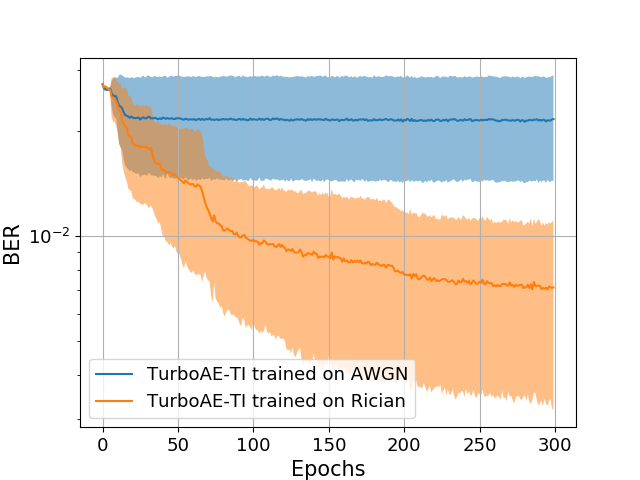}
\caption{Test BER along training epochs, averaged over 5 runs: Rician training leads to convergence to a lower BER.}\label{fig:riciantraining}
\end{figure}

\noindent\textbf{Effect of Rician training on \textsc{TurboAE-TI}. }\ To further understand the behavior of \textsc{TurboAE-TI} trained on Rician fading channels, in Figure~\ref{fig:riciantraining}, we plot the test BER performance on AWGN channel as a function of training epochs, comparing both training on Rician and AWGN channels when SNR = 1 dB. Rician training shows stable improvement while AWGN training tends to get stuck at local sub-optima. 

Given the general channel form as $\by = h\bx + \bz$, 
the gradient of loss $L(.)$ with respect to the encoder weight $\theta$ is:

\vspace{-1em}
\begin{align}
\label{enc_gradient}
    \frac{\partial L}{\partial \theta} = \frac{\partial L}{\partial \by} h \frac{\partial \bx}{\partial \theta}
\end{align}

When training with AWGN channel, the $h$ is constant. For Rician $K=10$, the realization of Rician fading channels is similar to the realization of AWGN channels, but the channel coefficient $h$ includes more perturbation. Based on this result, we conjecture that Rician training helps \textsc{TurboAE-TI} to escape the local sub-optima, via injecting gradient perturbations for training encoders.



\vspace{-1em}

\noindent\textbf{Partial Minimum Distance.\ }\
Additionally, we compute the minimum distance between codewords.  Since empirically computing minimum distance is intractable even for moderate block length, we propose an alternative method to evaluate the partial minimum distance. The notation is: $i$-th message $u_i$ of length $L=40$, with $M=10$, is composed as $u_i = [a, u_i^{M}, b]$, where $a$ and $b$ are the fixed random binary message of length $(L-M)/2$ across each enumerations. $u_i^{M}$ enumerates all possible message of length $M$. We enumerate all $2^M$ messages, and compute their partial minimum Euclidean distance:

\vspace{-.5em}
$$D_{min} = \min_{i \in \{1,..., 2^M\}, j \in \{1,..., 2^M\} , i \ne j} D(f(u_i), f(u_j))$$ 

\begin{figure}[!ht]
\centering
\includegraphics[width=0.7\linewidth]{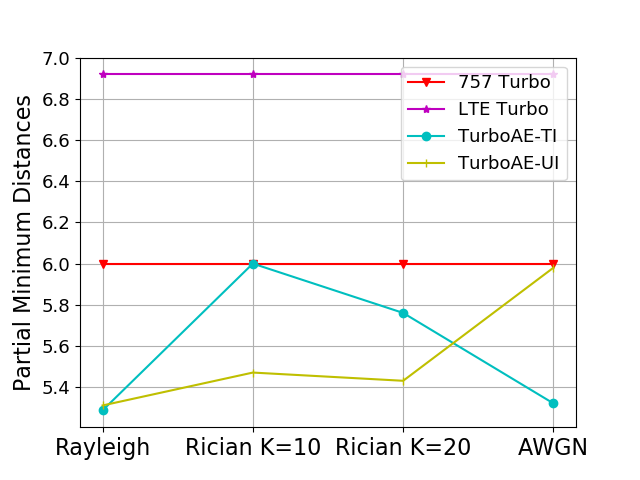}
\caption{Minimum distance of \textsc{TurboAE-TI} and \textsc{TurboAE-UI} trained on various channels: Rician training of \textsc{TurboAE-TI} leads to a larger minimum codeword distance than AWGN training. However, the minimum distance of \textsc{TurboAE-TI} is smaller than the one of LTE Turbo codes, which implies that 
\textsc{TurboAE-TI}'s superior performance is not fully captured in the distance.
}\label{fig:mindistance}
\end{figure}

The distance is also averaged over 100 instances. The results are shown in Figure~\ref{fig:mindistance}. For \textsc{TurboAE-TI}, Rician training with $K=10$ results in the best partial minimum distance, indicates that Rician training leads to better encoder. On the other hand, \textsc{TurboAE-UI} trained on Rician channel lead to worse performance. We conjecture that Rician training is more beneficial for \textsc{TurboAE-TI}, due to its sub-optima problem. Note that even the best \textsc{TurboAE-UI} partial minimum distance is still worse than regular Turbo code. We conjecture \textsc{TurboAE-UI}s learned sophisticated codes with patterns that are not fully captured in the codeword distance. On the other hand, this result shows that a further improvement might be feasible by deploying a regularizer on the codeword distance. 

\vspace{-1em}

\section{Conclusion} 


We introduce \textsc{TurboAE-TI}, where we make the interleaver of \textsc{TurboAE-UI} to be trainable by adding a matrix generalization of the $l_{1}-l_{2}$ penalty to our model's loss function, and introducing a well-thought-out training methodology.
We show that the trainable interleaver leads to the improved reliability on various channels, such as fading channels and bursty noise channels. The amount of gain varies from channel to channel and reaches up to 1dB. 

\textsc{TurboAE-TI} also outperforms \textsc{TurboAE-UI} for AWGN channels, when combined with the novel Rician training methodology. 
Our analysis suggests that the perturbation of channel weights in the Rician training helps the \textsc{TurboAE-TI} to escape the local optima. Moreover, Rician training also results in a better encoder, indicated by the improved partial minimum distance.

\section{Open problems} \label{sec:open}

There are several open problems that arise from this work. First, the extension of \textsc{TurboAE-TI} to \emph{multi-user scenarios} is an interesting future direction. In recent work~\cite{deepic2021}, it is shown that designing a code that utilizes interleavers for interference channels is challenging. With the aid of a trainable interleaver, we conjecture that one can learn a code with a long range memory for such \emph{multi-user scenarios} and hence achieve a better performance. 

 Moreover, developing deep learning methodologies driven by the communication theory is a promising direction. We conjecture that the Rician training framework can be applied to potentially enhance the performance of some state of the art applications (e.g., computer vision).



\begin{figure}[!ht]
    \centering
    \includegraphics[width=0.49\linewidth]{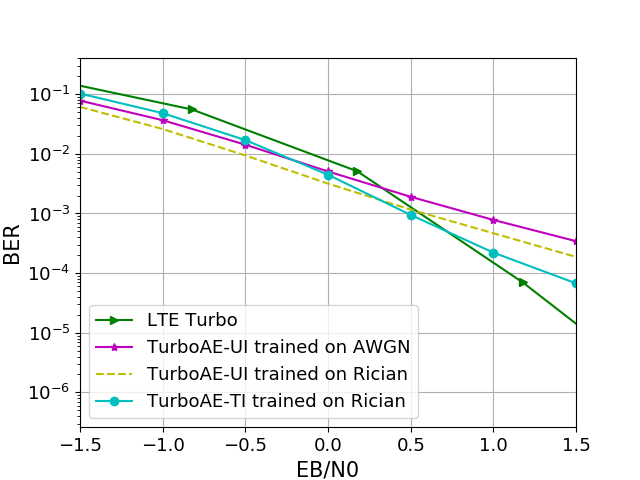}
    \includegraphics[width=0.49\linewidth]{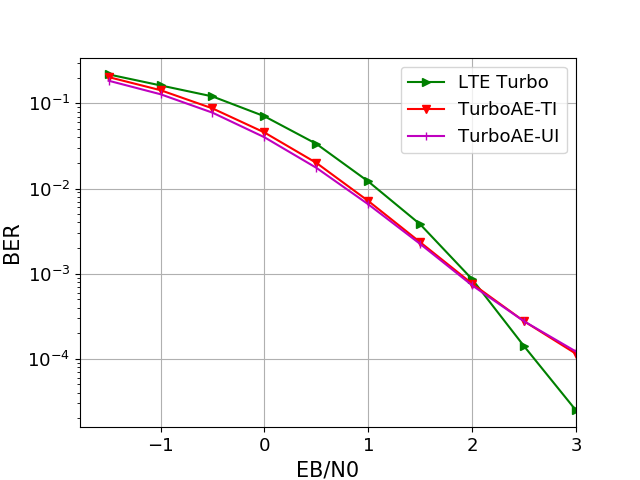}
    
    \caption{Results on L=100 AWGN (left) and Rician fading channels (right)}
    \label{fig:Turbosl100awgn}
\end{figure}

Finally, extending the \textsc{TurboAE-TI} framework to 
longer blocklengths and higher SNRs is challenging, as the interleaver's space grows in function of the blocklength, and the training becomes harder for higher SNRs. 
%
%
%
%
%
We illustrate our results on $L=100$ in Figure \ref{fig:Turbosl100awgn}. Although \textsc{TurboAE-TI} works better than \textsc{TurboAE-UI}, it performs worse than Turbo LTE. This indicates that the scalability of \textsc{TurboAE-TI} to a long blocklength is an interesting future research direction.

\vspace{-.8em} 
 \medskip
 \small
 \bibliographystyle{IEEEtran}
 \bibliography{References,megabib}

\newcommand{\noopsort}[1]{}
\begin{thebibliography}{10}
\providecommand{\url}[1]{#1}
\csname url@samestyle\endcsname
\providecommand{\newblock}{\relax}
\providecommand{\bibinfo}[2]{#2}
\providecommand{\BIBentrySTDinterwordspacing}{\spaceskip=0pt\relax}
\providecommand{\BIBentryALTinterwordstretchfactor}{4}
\providecommand{\BIBentryALTinterwordspacing}{\spaceskip=\fontdimen2\font plus
\BIBentryALTinterwordstretchfactor\fontdimen3\font minus
  \fontdimen4\font\relax}
\providecommand{\BIBforeignlanguage}[2]{{%
\expandafter\ifx\csname l@#1\endcsname\relax
\typeout{** WARNING: IEEEtran.bst: No hyphenation pattern has been}%
\typeout{** loaded for the language `#1'. Using the pattern for}%
\typeout{** the default language instead.}%
\else
\language=\csname l@#1\endcsname
\fi
#2}}
\providecommand{\BIBdecl}{\relax}
\BIBdecl

\bibitem{shannon1948mathematical}
C.~E. Shannon, ``A mathematical theory of communication, part i, part ii,''
  \emph{Bell Syst. Tech. J.}, vol.~27, pp. 623--656, 1948.

\bibitem{DBLP:journals/corr/OSheaH17}
T.~J. O'Shea and J.~Hoydis, ``An introduction to machine learning
  communications systems,'' \emph{CoRR}, vol. abs/1702.00832, 2017.

\bibitem{dorner2017deep}
S.~D{\"o}rner, S.~Cammerer, J.~Hoydis, and S.~t. Brink, ``Deep learning-based
  communication over the air,'' \emph{arXiv}, 2017.

\bibitem{gruber2017deep}
T.~Gruber, S.~Cammerer, J.~Hoydis, and S.~ten Brink, ``On deep learning-based
  channel decoding,'' in \emph{Information Sciences and Systems (CISS), 2017
  51st Annual Conference on}.\hskip 1em plus 0.5em minus 0.4em\relax IEEE,
  2017, pp. 1--6.

\bibitem{alkhateeb2018deep}
A.~Alkhateeb, S.~Alex, P.~Varkey, Y.~Li, Q.~Qu, and D.~Tujkovic, ``Deep
  learning coordinated beamforming for highly-mobile millimeter wave systems,''
  \emph{IEEE Access}, vol.~6, pp. 37\,328--37\,348, 2018.

\bibitem{modulation1}
S.~Park, H.~Jang, O.~Simeone, and J.~Kang, ``Learning how to demodulate from
  few pilots via meta-learning,'' in \emph{2019 IEEE SPAWC}, 2019, pp. 1--5.

\bibitem{ye2017power}
H.~Ye, G.~Y. Li, and B.-H. Juang, ``Power of deep learning for channel
  estimation and signal detection in ofdm systems,'' \emph{IEEE Wireless
  Communications Letters}, vol.~7, no.~1, pp. 114--117, 2017.

\bibitem{csinet2}
T.~Wang, C.-K. Wen, S.~Jin, and G.~Y. Li, ``Deep learning-based csi feedback
  approach for time-varying massive mimo channels,'' \emph{IEEE Wireless
  Communications Letters}, vol.~8, no.~2, pp. 416--419, 2019.

\bibitem{nachmani2018deep}
E.~Nachmani, E.~Marciano, L.~Lugosch, W.~J. Gross, D.~Burshtein, and
  Y.~Be’ery, ``Deep learning methods for improved decoding of linear codes,''
  \emph{IEEE Journal of Selected Topics in Signal Processing}, vol.~12, no.~1,
  pp. 119--131, 2018.

\bibitem{shlezinger2019viterbinet}
N.~Shlezinger, N.~Farsad, Y.~C. Eldar, and A.~J. Goldsmith, ``Viterbinet:
  Symbol detection using a deep learning based viterbi algorithm,'' \emph{IEEE
  20th International Workshop on Signal Processing Advances in Wireless
  Communications (SPAWC)}, 2019.

\bibitem{turbonet20}
Y.~He, J.~Zhang, S.~Jin, C.-K. Wen, and G.~Y. Li, ``Model-driven {DNN} decoder
  for turbo codes: Design, simulation, and experimental results,'' \emph{IEEE
  Transactions on Communications}, vol.~68, no.~10, 2020.

\bibitem{Jiang19}
Y.~Jiang, H.~Kim, H.~Asnani, S.~Kannan, S.~Oh, and P.~Viswanath, ``Turbo
  autoencoder: Deep learning based channel code for point-to-point
  communication channels,'' in \emph{Advances in Neural Information Processing
  Systems (NeurIPS)}, 2019.

\bibitem{kim2018deepcode}
H.~Kim, Y.~Jiang, S.~Kannan, S.~Oh, and P.~Viswanath, ``Deepcode: Feedback
  codes via deep learning,'' in \emph{Advances in Neural Information Processing
  Systems (NeurIPS)}, 2018, pp. 9436--9446.

\bibitem{serialturboae}
J.~Clausius, S.~D{\"{o}}rner, S.~Cammerer, and S.~ten Brink, ``Serial vs.
  parallel turbo-autoencoders and accelerated training for learned channel
  codes,'' vol. abs/2104.14234, 2021.

\bibitem{pmlr-v139-makkuva21a}
A.~V. Makkuva, X.~Liu, M.~V. Jamali, H.~Mahdavifar, S.~Oh, and P.~Viswanath,
  ``Ko codes: inventing nonlinear encoding and decoding for reliable wireless
  communication via deep-learning,'' in \emph{Proceedings of the 38th
  International Conference on Machine Learning (ICML)}, 2021.

\bibitem{ftae2020}
Y.~Jiang, H.~Kim, H.~Asnani, S.~Oh, S.~Kannan, and P.~Viswanath, ``Feedback
  {T}urbo autoencoder,'' in \emph{2020 IEEE ICASSP}, 2020.

\bibitem{turboaemod}
Y.~Jiang, H.~Kim, H.~Asnani, S.~Kannan, S.~Oh, and P.~Viswanath, ``Joint
  channel coding and modulation via deep learning,'' in \emph{2020 IEEE SPAWC},
  2020, pp. 1--5.

\bibitem{esser2013method}
E.~Esser, Y.~Lou, and J.~Xin, ``A method for finding structured sparse
  solutions to nonnegative least squares problems with applications,''
  \emph{SIAM Journal on Imaging Sciences}, vol.~6, no.~4, pp. 2010--2046, 2013.

\bibitem{deepic2021}
H.~K. Karl~Chahine, Nanyang~Ye, ``{D}eep{IC}: Coding for interference channels
  via deep learning,'' \emph{GlobeCom}, 2021.

\end{thebibliography}

\end{document}